\documentstyle[11pt,aaspp4]{article}
\slugcomment{\shortstack[r]{To appear in ApJL}}
\def\lsim{\mathrel{\mathpalette\Oversim<}}
\def\gsim{\mathrel{\mathpalette\Oversim>}}
\def\Oversim#1#2{\lower0.5ex\vbox{\baselineskip0pt\lineskip0pt%
            \lineskiplimit0pt\ialign{%
          $\mathsurround0pt #1\hfil##\hfil$\crcr#2\crcr\sim\crcr}}}

\begin{document}
\newcommand{\tcole}{t_{\rm cool,H_2}^{\rm eq}}
\newcommand{\tcoles}{t_{\rm cool}^{\rm eq}}
\newcommand{\tcolo}{t_{\rm cool,oth}}
\newcommand{\tcol}{t_{\rm cool}}
\newcommand{\tdis}{t_{\rm dis}}
\newcommand{\trec}{t_{\rm rec}}
\newcommand{\ymoe}{y_{\rm H_2}^{\rm eq}}
\newcommand{\ymo}{y_{\rm H_2}}

\title{Formation and Disruption of Cosmological Low Mass Objects}
\author{Ryoichi Nishi\altaffilmark{1} 
\affil{Department of Physics, Kyoto University, Kyoto 606-8502, Japan}
and \\
Hajime Susa\altaffilmark{2}
\affil{Center for Computational Physics, University of
  Tsukuba, Tsukuba 305-8571, Japan }}
\altaffiltext{1}{e-mail:nishi@tap.scphys.kyoto-u.ac.jp}
\altaffiltext{2}{e-mail:susa@rccp.tsukuba.ac.jp}

\abstract{
We investigate the evolution of cosmological low mass 
(low virial temperature) objects and 
the formation of the first luminous objects. First, the `cooling diagram' 
for low mass objects is shown. 
We assess the cooling rate taking into account the 
contribution of H$_2$, which is not in chemical equilibrium generally, 
with a simple argument of time scales. 
The reaction rates and the cooling rate of H$_2$ are taken from 
the recent results by Galli \& Palla (1998).
Using this cooling diagram, 
we also estimate the formation condition of luminous objects 
taking into account the supernova (SN) disruption of virialized clouds. 
We find that the mass of the first luminous object is several $\times
10^7 M_\odot$, because smaller objects may be disrupted by the SNe 
before they become luminous.  
Metal pollution of low mass (Ly-$\alpha$) clouds also discussed.
The resultant metallicity of the clouds is $Z/Z_\odot \sim 10^{-3}$.
}


\newpage
\section{Introduction}
Today, we have a great deal of observational data concerning the 
early universe. 
However, we have very little information about the era referred to as 
the `dark ages'.
Information regarding the era of recombination 
(with redshift $z$ of about $10^3$) can be obtained 
by the observation of cosmic microwave background radiation. 
After the recombination era little information is accessible 
until $z \sim 5$, 
after that we can observe luminous objects such as galaxies and QSOs. 
On the other hand, the reionization of the intergalactic medium and 
the presence of heavy elements at high-$z$  
suggest that there are other population of luminous objects, 
which precedes normal galaxies.
Thus, theoretical approach to reveal the formation mechanism of 
the such unseen luminous objects is very important.

It is now widely accepted that luminous objects 
are formed from overdense regions in the early universe. 
These overdense regions collapse to form luminous objects, 
in case they fragment into {\it many} stellar size clouds 
and {\it many} massive stars are formed.
In order to understand the way in which luminous objects are formed, 
physical processes of the clouds in various stages of 
evolution should be studied, individually.

The formation process of luminous objects is roughly divided into 
three steps, formation of cold clouds by H and/or ${\rm H_2}$ line 
cooling, formation of the first generation stars in the cold clouds, 
and the star formation throughout the host clouds. These steps are 
disturbed by the feedback from the first stars. 
The first step has been investigated by many authors 
(e.g., \cite{HTL96,OG96,Tegmark97,GO97,Abel98}) and it has been shown 
that the low mass clouds (virial temperature is several 
$\times 10^3$ K) become the earliest cooled dense clouds. 
The second step, however, is not investigated enough, although 
initial mass function and formation efficiency of the first generation stars 
in the clouds are very challenging and crucial problems.
Many authors attacked this problem (\cite{MST69,Hut76,Carl81,PSS83,Uehara96}) 
and they obtained various conclusions. 
However, now, the mass of the first generation stars 
are estimated through detailed investigation to be fairly large 
(\cite{NU98,ON98}). 
For the third step, the feedback from the luminous objects on the 
other clouds has been 
studied by several authors (\cite{HRL96,HRL97,Ferr98,HAR99}).
Haiman, Abel \& Rees (1999) examined the build-up of the UV background 
in hierarchical models and its effects on star formation inside small 
halos that collapse prior to reionization. They stressed that early UV 
background below 13.6 eV suppresses the H$_2$ abundance and there 
exists a negative feedback even before reionization. 
Moreover, the feedback from the formed stars on their own host cloud is more 
serious. 
The main feedback consists of two
different processes, UV radiation from the stars and energy input by SNe.
Through ionization of H (\cite{LM92}) and dissociation of H$_2$ 
(\cite{Silk77,ON99}), UV radiation have negative feedback on the further 
star formation in the host clouds. Especially, H$_2$ is dissociated in such 
a large region that the whole of an ordinary low mass cloud 
is influenced by one O5 type star (\cite{ON99}). 
The feedback from SNe on the host clouds 
is probably negative (e.g., \cite{MF99}). SNe can disrupt the host
clouds before they become luminous, because the explosion energy is
comparable with the typical binding energy of host clouds. 
In this {\it Letter}, 
we investigate the evolution of low mass primordial clouds 
systematically, and assess the mass of the first luminous objects. 

\section{Cooling diagram}
The formation of cold dense clouds, i.e., progenitors of luminous objects, 
is basically understood by the comparison between free-fall time and the
cooling time. The `cooling diagram' originally introduced by Rees \&
Ostriker (1977) and 
Silk (1977) shows the region where cooling time is shorter than the
free-fall time on $\rho$ - $T$ plane, and vice versa.
In this section, we present the cooling diagram on $\rho$ - $T$ plane
including H$_2$ cooling. 
With this diagram, 
we can predict whether a cloud virialized at $z=z_{\rm vir}$ with
virial temperature $T_{\rm vir}$ cools.
\subsection{H$_2$ fraction with given virial temperature}
\begin{figure}
\plotone{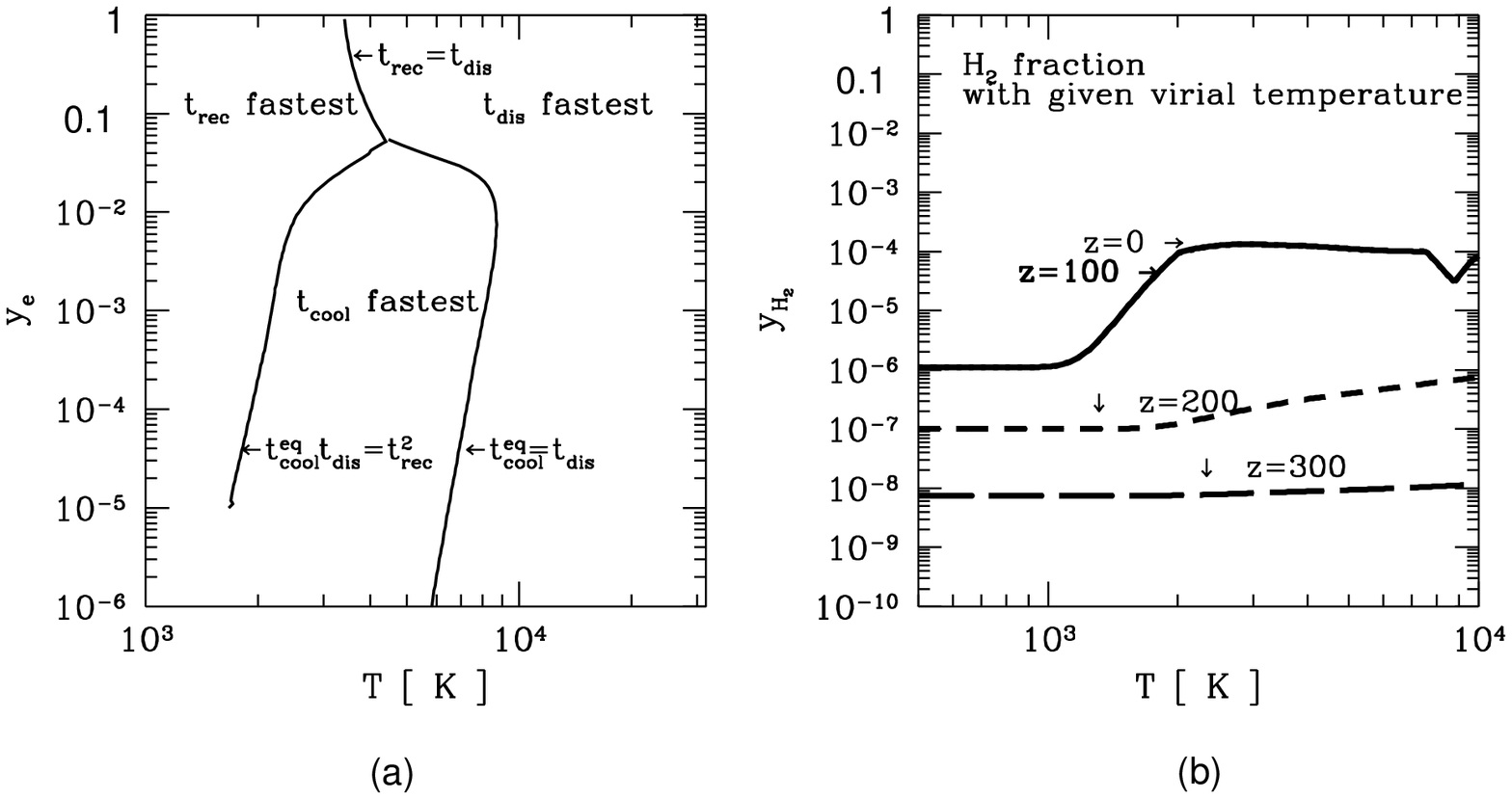}
\caption[dummy]{(a) The $y_e$ - $T$ plane is divided into three region. 
In the region with higher temperature, $t_{\rm dis}$ is the fastest and 
$y_{\rm H_2}=y_{\rm H_2}^{\rm eq}$. In the region with lower temperature, 
$t_{\rm rec}$ is the fastest and 
$y_{\rm H_2}=y_{\rm H_2}^{\rm eq} {t_{\rm rec} \over t_{\rm dis}}$.
Between these two regions, where $t_{\rm cool}$ is the fastest, 
$y_{\rm H_2} \simeq \ymoe\sqrt{\frac{\tcoles}{\tdis}}$.
(b) Virial temperature v.s. H$_2$ fraction is plotted for given
 epoch of collapse ($z$).
}
\label{fig1}
\end{figure}

In order to estimate the cooling rate at $T<10^4$K, we need the
fraction of H$_2$. 
The number fraction of H$_2$ (hereafter denoted as $\ymo$) is not 
generally in 
equilibrium for $T < 10^4$ K in the epoch of galaxy formation.
The value of $\ymo$ at a time depends not only on $\rho$ and $T$ but also 
on the initial condition. 
Consequently, we cannot evaluate the cooling rate on the $\rho$ - $T$
plane without estimating non-equilibrium $\ymo$. 
Tegmark et al. (1997) calculate $\ymo$ numerically, however, their 
primordial $\ymo$ is about two orders of magnitude over estimated because 
the destruction rate of ${\rm H_2^+}$ by cosmic microwave background 
radiation at high-$z$ is under estimated (\cite{GP98}). 
Since their primordial value 
$\sim 10^{-4}$ is comparable to the necessary value to cool, 
their cooling criterion is not reliable generally.  
Here, using recent reaction rates and the cooling rate of H$_2$ 
(\cite{GP98}), 
we adopt a simplified and generalized method to estimate the
cooling function of H$_2$, differently from that of Tegmark et al. (1997). 
We introduce four important time scales, 
$\tdis$, $t_{\rm form},\tcol,$ and $\trec$. They represent
 dissociation and formation time of H$_2$, cooling time, and
recombination time, respectively. Comparing these time scales, we
assess the non-equilibrium fraction of H$_2$ with given virial temperature, 
and redshift.
Below, our estimation is summarized (see Fig. \ref{fig1}a).

1. The case $\tdis < {\rm min}( \tcol, \trec)$ (Region of 
``$t_{\rm dis}$ fastest'' in Fig. \ref{fig1}a): 
H$_2$ is in chemical equilibrium. 
In this case, $\ymo=\ymoe$, where $\ymoe$ denotes the fraction of H$_2$ 
in chemical equilibrium (solution of $t_{\rm form}=\tdis$). 

2. The case $\trec< {\rm min}( \tcol, \tdis)$ 
(Region of ``$t_{\rm rec}$ fastest'' in Fig. \ref{fig1}a): 
H$_2$ is out of  chemical equilibrium, and H$_2$ molecules are formed 
until the recombination process significantly reduces the electron fraction. 
As a result, $\ymo$ is determined by the equation, $t_{\rm form}=\trec$. 
Combined with the relation $t_{\rm form}=\ymo/\ymoe\tdis$(\cite{Susa98}),
$\ymo$ is obtained as,
$\ymo=\ymoe\left(\trec/\tdis\right)$.

3. The case $\tcol < {\rm min}(\trec, \tdis)$ 
(Region of ``$t_{\rm cool}$ fastest'' in Fig. \ref{fig1}a): 
When the cooling time is the shortest of the three time scales, $\ymo$ is
determined by the equation $t_{\rm form}=\tcol$. In other words,
$\ymo$ increases until the system is cooled
significantly. 
In case $\ymo$ cooling dominates the other cooling processes, 
$\ymo\simeq\ymoe\sqrt{\tcoles/\tdis}$. Otherwise, $\ymo$ is the solution
of a quadratic equation.
Here, $\tcoles$ represents cooling time scale by H$_2$ rovibrational
transitions with $\ymoe$.

The electron fraction of a virialized cloud is assumed as
$y_e={\rm max}(y_e^{\rm rel},y_e^{\rm eq})$.
Here $y_e^{\rm rel}$ is the fraction of cosmologically relic
electrons calculated in Galli \& Palla (1998). It equals to $3.02\times
10^{-4}$ for their standard model.
The chemical equilibrium fraction of electrons is denoted $y_e^{\rm eq}$. 
With this electron fraction, we estimate the fraction of H$_2$.

In Fig. \ref{fig1}b, $\ymo$ is plotted with given virial temperatures for
four redshifts. For low redshift ($z \lsim 100$) and high temperature
($T\simeq 10^4 $K), H$_2$ is in chemical equilibrium with given
ionization degree. 
As the temperature drops, H$_2$ gets out of equilibrium because the cooling
time becomes shorter than the other time scales. Below $2000$ K,
recombination time scale is the shortest, and $\ymo$ becomes relic value.

For high redshift, the destruction of H$^-$ ($z \gsim 100$) and
H$^+_2$ ($z \gsim 200$) by cosmic microwave background radiation reduces 
$\ymo$ significantly.

\subsection{Comparison between free-fall time and cooling time}
We are able to assess the cooling rate with the given H$_2$ fraction 
evaluated in the previous subsection.
We compare the time scale of collapse ($t_{\rm ff}$) with the
cooling time ($t_{\rm cool}$) which include the contribution from the
H$_2$ cooling.
They are, 

\begin{equation}
t_{\rm ff}=\left(\frac{3\pi}{32G\rho_{\rm vir}}\right)^{1/2}, 
~~~t_{\rm cool}=\frac{1.5 \mu^{-1} k T_{\rm vir}} 
{n_{\rm vir}\Lambda\left(\ymo,T_{\rm vir},n_{\rm vir}\right)}.
\label{eq:tfc}
\end{equation}
Here, $\rho_{\rm vir} \equiv 18 \pi^2 \Omega \rho_{\rm cr}$ and 
$n_{\rm vir} \equiv \Omega_b \rho_{\rm vir} / m_p$, where 
$\rho_{\rm cr} \equiv 1.9 \times 10^{-29} h^2 (1 + z_{\rm vir})^3 
{\rm ~g ~cm}^{-3}$. 
We adopt $\Omega=1$, $\Omega_b=0.06$ and $h=0.5$ in this paper.

Equating $t_{\rm ff}$ and $t_{\rm cool}$ in eq. 
(\ref{eq:tfc}), we obtain the boundary between the cooled region
during the collapse and the other region, 
which is drawn on the $(1+z_{\rm vir})$ - $T_{\rm vir}$ plane in
Fig. \ref{fig2}a. The objects virialized into the region denoted as
$t_{\rm ff} < t_{\rm cool}$ will be cooled by H$_2$ during the
gravitational collapse. In this case, collapsing cloud will be a
mini-pancake, because the thermal pressure becomes negligible.
We remark that the cooling region expands into $T_{\rm vir}\lsim 10^4$K,
which is different from classical cooling diagram such as the one 
in Rees \& Ostriker (1977).

We also compare the cooling time scale with the Hubble expansion
time ($H^{-1}$).
The line $t_{\rm cool}=H^{-1}$ is also drawn on
Fig. \ref{fig2}a. The cooling region during the Hubble expansion time is
slightly larger than the previous one, because Hubble expansion time is
longer than the free-fall time. In this case, the collapse proceeds in
semi-statically. As a result, the central region of the cloud will
proceed to the run away collapse phase (\cite{TI99}).

\section{SNe and disruption of the bound objects}
As the collapse proceeds, small amount of the total gas is cooled to
$100$K by H$_2$.
In those clouds, massive stars ($10-100
M_\odot$) will be formed (\cite{NU98,ON98}), eventually. 
After massive first generation stars form, evolution of the host 
clouds become slower because of strong regulation by UV radiation 
(\cite{ON98}). 
Thus, next generation stars are hardly formed before the first generation 
stars die. 
Subsequent SNe might disrupt the gas binding before significant amount
of total gas transferred into stars. 
Here, we derive the cloud disruption condition by SNe,  
with the assumption that the cloud is spherical and the density 
is constant, for the simplicity.

We estimate the kinetic energy transferred from the SNe to gas.
The velocity of expanding shock front from the center of a supernova 
remnant (SNR) is
\begin{eqnarray}
v_{\rm s}(t)=\left(\frac{7.64\times 10^{-3}\left(\gamma^2-1\right)KE}
{\rho_1}\right)^{1/5}t^{-3/5}, \label{eq:shockv}
\end{eqnarray}
where $K=1.53$, $E$ is the total thermal energy given by the SN, $\rho_1$
is the density of the cloud before the explosion, and $t$ denotes the
elapsed time since the explosion (\cite{Spitzer78}).
Integrating eq. (\ref{eq:shockv}), we obtain the location of the
shock front:
\begin{eqnarray}
R_{\rm s}(t)=\left(\frac{0.746\left(\gamma^2-1\right)KE}
{\rho_1}\right)^{1/5}t^{2/5}. \label{eq:shockr}
\end{eqnarray}
The mass of the hot bubble is also obtained as 
$m_{\rm SNR}=\frac{4}{3}\pi R_{\rm s}^3(t) \rho_1$.
The hot bubble keeps pushing the surrounding gas until the thermal
energy is pumped off by the radiative cooling.
Thus, the total momentum transferred from the SNR to the gas cloud is
$p_{\rm tot}= m_{\rm SNR}(t_{\rm cool}) 
v_{\rm s}(t_{\rm cool})$. 
Equating this momentum with the momentum of the whole cloud, we have the
expanding velocity of the cloud:
\begin{eqnarray}
v_{\rm tot}=\frac{m_{\rm SNR}\left(t_{\rm cool}\right)}{m_{\rm tot}}
v_s\left(t_{\rm cool}\right).\label{eq:vtot}
\end{eqnarray} 

If this velocity $v_{\rm tot}$ is smaller than the escape velocity
($v_{\rm esc}$) of
the cloud, it will be still bounded. Otherwise, it disrupts.
Now, we replace $m_{\rm tot}$ in equation (\ref{eq:vtot}) 
with $m_{\rm J}(T_{\rm vir},z_{\rm vir})$,
which is the virialized mass of the cloud collapsed at $z_{\rm vir}$ with
$T_{\rm vir}$. The escape velocity from the cloud is directly related to the
virial temperature. Consequently, we can draw the disruption boundary 
($v_{\rm esc}=v_{\rm tot}$) on the $(1+z_{\rm vir})$ - $T_{\rm vir}$ plane.

On the cooling diagram (Fig. \ref{fig2}a), the boundaries ($v_{\rm
esc}=v_{\rm tot}$) are superimposed for two cases. These lines are 
obtained with the assumption that the input thermal energy from the SNe
is $10^{51}$ erg and $10^{52}$ erg, respectively. 
The input energy almost reflects the number of the SNe. The values 
$10^{51}$ erg and $10^{52}$ erg represent the case of single SN 
and $10$ SNe, respectively.
The former should corresponds to the clouds in $t_{\rm ff}<t_{\rm
cool}<H^{-1}$, because they will have a runaway collapsing central
core. The core evolves much faster than the envelope 
and will be a massive star, probably, followed by a single SN. 
The latter case
represents the clouds in $t_{\rm ff}>t_{\rm cool}$. They will have a
shocked pancake, and the cooled region will fragment into
stars. That's why they should have multiple SNe. 

However, we should note that the disruption criteria by SNe 
strongly depend on the geometry of objects (\cite{MF99,Ciardi99} and 
references there in). 
In the case that cooling is efficient ($t_{\rm ff}>t_{\rm cool}$), 
a cloud evolves dynamically and becomes complicated shape, which is
probably flattened. The geometric effect makes the momentum transfer 
from the SNR to the surrounding gas less efficient than our evaluation. 
On the other hand, if cooling is not efficient ($t_{\rm ff}<t_{\rm cool}$), 
a cloud becomes fairly spherical and has a centrally condensed density
profile. 
In this case, the effect of SNe may become stronger than 
the above estimate (e.g., \cite{ML89}). 
Moreover, if the duration of the multiple SNe is longer than  
or comparable with the evolution time scale of a SNR, 
disruption criteria is not evaluated only with the total energy 
of multiple SNe (e.g., \cite{CF97}). 
Thus, our estimate shows the qualitative tendency, so that 
detailed calculation for a individual cloud is necessary to 
derive the SNe effects accurately. 

\section{Evolution of low mass objects and mass of the first luminous objects}
\begin{figure}
\plotone{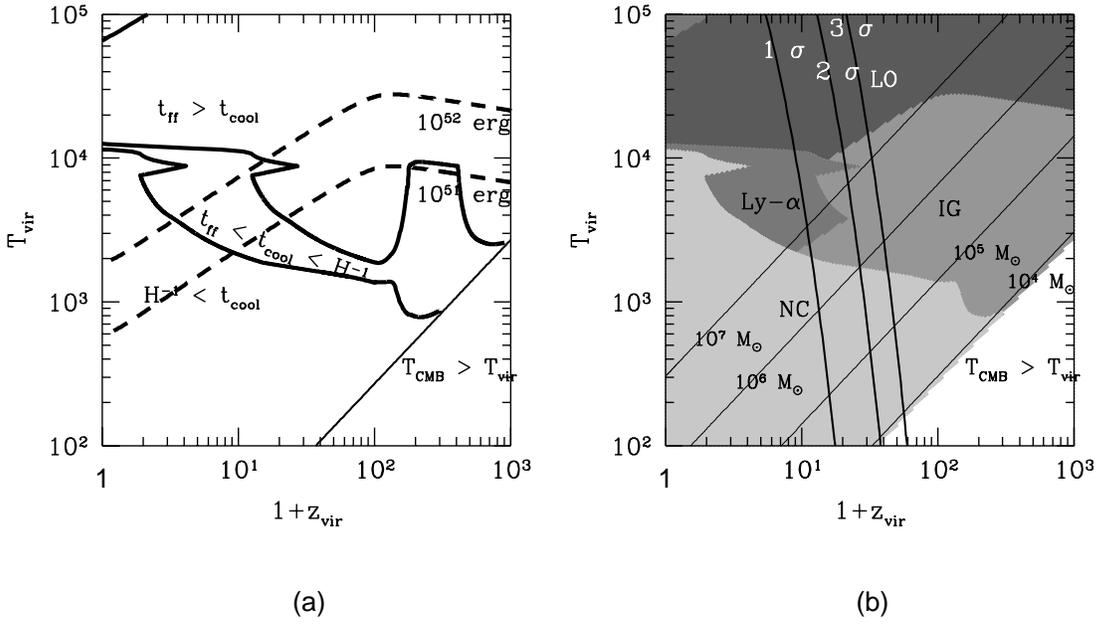}
\caption[dummy]{(a) Cooling diagram is plotted on the $(1+z_{\rm vir})$ -
 $T_{\rm vir}$ plane. The thick solid lines divide the plane into three
 regions. In the region denoted as $t_{\rm ff} > t_{\rm cool}$, cooling
 proceeds faster than the collapse. The narrow region denoted as 
$t_{\rm ff} < t_{\rm cool} < H^{-1}$, the cloud cools 
faster than the Hubble expansion, but cannot catch up with the
 free-fall.
The clouds virialized into the lower region ($H^{-1} < t_{\rm cool}$), 
cannot cool and will not be the luminous objects.
The lower right region ($T_{\rm CMB} > T_{\rm vir}$) is the forbidden
 region of cooling, due to the Compton heating.
Two dashed lines denote the boundary below which virialized clouds are
 destroyed by SN explosions. The lines correspond to the cases that
 the total energy of SNe is $10^{51}$ erg and  $10^{52}$ erg.
(b) $(1+z_{\rm vir})$ - $T_{\rm vir}$ plane is divided into five
 regions. In the unshaded lower right region ($T_{\rm CMB} > T_{\rm vir}$)
 and the lightly shaded region denoted as ``NC'', 
clouds are diffuse and do not become luminous, 
 because radiative cooling is not efficient. In the shaded
 region denoted as ``IG'', clouds will be disrupted by the SNe. 
Clouds in the region ``Ly-$\alpha$'' will not be
 destroyed by the SN. 
The dark shaded upper region (denoted as ``LO''),
 represents the clouds which could be luminous objects. 
The thick solid three lines marked as $1\sigma$, $2\sigma$, and
 $3\sigma$ represent the location that the density perturbations in
 standard CDM cosmology should virialize. 
The mass of the clouds are constant along the thin solid four lines 
which cross the panel from lower left to upper right. 
The lines correspond to the total masses are  
$10^4 M_\odot$, $10^5 M_\odot$, $10^6 M_\odot$, and $10^7 M_\odot$.
}

\label{fig2}
\end{figure}

According to the argument in the previous section, 
the clouds in the region 
$t_{\rm ff}>t_{\rm cool}$ will experience multiple SNe. 
Assuming that the total energy of multiple SNe as $10^{52}$ erg, 
the survived region is the dark shaded upper region of 
Fig. \ref{fig2}b (denoted as ``LO'').\footnote{Of course there 
exists some ambiguity in the total energy, but Fig. \ref{fig2}b dose not 
change by this ambiguity qualitatively.} 
As shown in Fig. \ref{fig2}b survived clouds are fairly 
massive ($T_{\rm vir} \gsim 10^4$ K) 
and they evolve into luminous objects 
through following processes (\cite{Nishi98}):  
(1) By pancake collapse of an overdense region 
or collision between subclouds in a potential well, 
a quasi-plane shock forms (e.g., \cite{SUN96}). 
(2) If the shock-heated temperature is higher than $\sim 10^4$~K, 
the post-shock gas is ionized and cooled efficiently by H line cooling. 
After it is cooled bellow $10^4$~K, ${\rm H_2}$ is formed fairly 
efficiently and it is cooled to several  hundred K by 
${\rm H_2}$ line cooling (e.g., \cite{SK87,Susa98}).  
(3) The shock-compressed layer fragments into cylindrical clouds 
when $t_{\rm dyn} \sim t_{\rm frag}$ (\cite{YN98,UN99}). 
(4) The cylindrical cloud collapses dynamically 
and fragments into cloud cores when $t_{\rm dyn} \sim t_{\rm frag}$ 
(\cite{Uehara96,NU98}). 
(5) Primordial stars form in cloud cores 
(\cite{ON98}). 
(6) Since the gravitational potential of the cloud is deep enough, 
subsequent SNe cannot disrupt the cloud. Star formation regulation 
by UV radiation is 
also weak because of highly flattened configuration of the host cloud. 
(7) Next generation stars can form efficiently and 
the cloud evolves into a luminous object. 

On the other hand, the clouds in the 
region $t_{\rm ff}<t_{\rm cool}<H^{-1}$ will experience a single SN.
As a result, the survived region is bounded by the line denoted as
$10^{51}$ erg (denoted as ``Ly-$\alpha$'' of Fig. \ref{fig2}b) 
and they evolve into luminous objects if they are isolated.
However, the evolution time scale of these clouds is rather long 
because of large disturbance by the SN and they are reionized at low-$z$.
\footnote{If the first star dies without SN and becomes 
black hole, metal pollution of the cloud does not occur. 
However, considering the star formation regulation by 
UV radiation,\cite{ON99}  
the evolution of the cloud is still slow and it is not likely to evolve 
into luminous object.}
After reionization, they may be observed as Ly-$\alpha$ clouds. 
Since the baryonic mass of these clouds are several 
$\times 10^5 M_\odot$ and the ejected metal mass by a SN is 
several $M_\odot$, their metallicity is estimated to be 
$\sim 10^{-3} Z_\odot$. 
The observation of QSO absorption line systems imply the
similar metallicity for Ly-$\alpha$ clouds 
(\cite{Cow95}; \cite{SC96}; \cite{Son97}; \cite{CS98}).
They are typically the 1$\sigma$ objects and 
collapse at $z \sim 10$. 
Our estimate basically agrees with 
the more detailed calculation in Ciardi \& Ferrara (1997).  

In the unshaded lower right region ($T_{\rm CMB} > T_{\rm vir}$)
 and the lightly shaded region ``NC'' of Fig. \ref{fig2}b, 
clouds are diffuse and do not become luminous, 
 because radiative cooling is not efficient. In the shaded
 region ``IG'', SNe destroy the binding of host objects, followed by the
diffusion of heavy elements into the surrounding medium.

Therefore, the first luminous objects are probably formed 
in the region ``LO'' 
and their mass is estimated to be several 
$\times 10^7 M_\odot$, if we consider the $2 \sim 3 \sigma$ 
objects. Formation epoch of the first luminous objects is 
$z\sim 30$ (considering the 3$\sigma$ objects) or $z\sim 20$ 
(considering the 2$\sigma$ objects). 
This estimated mass is larger than 
the one obtained by Tegmark et al. (1997), 
because small clouds may be blown up by their own SNe.

\acknowledgments
We would like to thank the anonymous referee for valuable comments. 
This work is supported by Research Fellowships of the Japan Society 
for the Promotion of Science for Young Scientists, No.2370 (HS) 
and by the Japanese Grant-in-Aid 
for Scientific Research on Priority Areas (No. 10147105) (RN) and 
Grant-in-Aid for Scientific Research of the Ministry of Education,
Science, Sports and Culture of Japan, No. 08740170 (RN). 

\newpage

\newpage
\end{document}